\documentclass[sigconf]{acmart}
\usepackage{amsmath,amssymb,amsfonts}
\usepackage{algorithmic}
\usepackage{graphicx}
\usepackage{textcomp}
\usepackage{rotating}
\usepackage{multirow}

\AtBeginDocument{%
  \providecommand\BibTeX{{%
    \normalfont B\kern-0.5em{\scshape i\kern-0.25em b}\kern-0.8em\TeX}}}

\setcopyright{acmcopyright}
\copyrightyear{2020}
\acmYear{2020}
\acmDOI{}

\acmConference[ICSE 20]{ICSE \'20: International Conference on Software Engineering}{May 23--29, 2020}{Seoul, South Korea}
\acmBooktitle{}
\acmPrice{}
\acmISBN{}

\begin{document}

\title{The SmartSHARK Ecosystem for Software Repository Mining}%

\author{Alexander Trautsch}
%\orcid{}
\email{alexander.trautsch@cs.uni-goettingen.de}
\affiliation{%
  \institution{Institute of Computer Science\\University of Goettingen}
  \country{Germany}
}

\author{Fabian Trautsch}
%\orcid{}
\email{fabian.trautsch@cs.uni-goettingen.de}
\affiliation{%
  \institution{Institute of Computer Science\\University of Goettingen}
  \country{Germany}
}

\author{Steffen Herbold}
\email{steffen.herbold@cs.uni-goettingen.de}
\affiliation{%
  \institution{Institute of Computer Science\\University of Goettingen}
  \country{Germany}
}

\author{Benjamin Ledel}
\email{benjamin.ledel@stud.uni-goettingen.de}
\affiliation{%
  \institution{Institute of Computer Science\\University of Goettingen}
  \country{Germany}
}

\author{Jens Grabowski}
\email{jens.grabowski@cs.uni-goettingen.de}
\affiliation{%
  \institution{Institute of Computer Science\\University of Goettingen}
  \country{Germany}
}

\begin{abstract}
Software repository mining is the foundation for many empirical software engineering studies. The collection and analysis of detailed data can be challenging, especially if data shall be shared to enable replicable research and open science practices. SmartSHARK is an ecosystem that supports replicable and reproducible research based on software repository mining.
\end{abstract}



\maketitle

\newcommand{\etal}{~\textit{et al.}}

\section{Introduction}

Mining software repositories (MSR) has become a standard technique that is frequently employed for empirical software engineering. MSR relies on tools that can extract data from repositories such as version control systems like Git or SVN, issue tracking systems like Jira or Bugzilla, or question/answer sites like StackOverflow. 

With this demonstration, we show the capabilities of Smart\-SHARK, an ecosystem for reproducible mining of software repositories~\cite{Trautsch2017}. SmartSHARK is designed with the goal to support replications in the context of MSR. The concept of SmartSHARK is to collect data from different sources and store all data in a single database with a harmonized schema. Analysis approaches can then rely on the harmonized data representation, which simplifies working with different data sources, allows automated re-runs of experiments on newly collected data, and facilitates the creation of benchmarks with a common interface. 

The remainder of this paper is structured as follows. We start with an overview of the SmartSHARK ecosystem in Section~\ref{sec:ecosystem}, describe the data that can currently be collected with SmartSHARK,  and describe the commonly used database, as well as components that support data collection, visulization, and validation. Then, we describe how SmartSHARK can be used for the analysis of the collected data in Section~\ref{sec:analysis}. Section~\ref{sec:related-work} discusses related work. Finally, we conclude in Section~\ref{sec:conclusion}.

\section{The SmartSHARK Ecosystem}
\label{sec:ecosystem}

\begin{figure}
    \centering
    \includegraphics[width=0.9\linewidth]{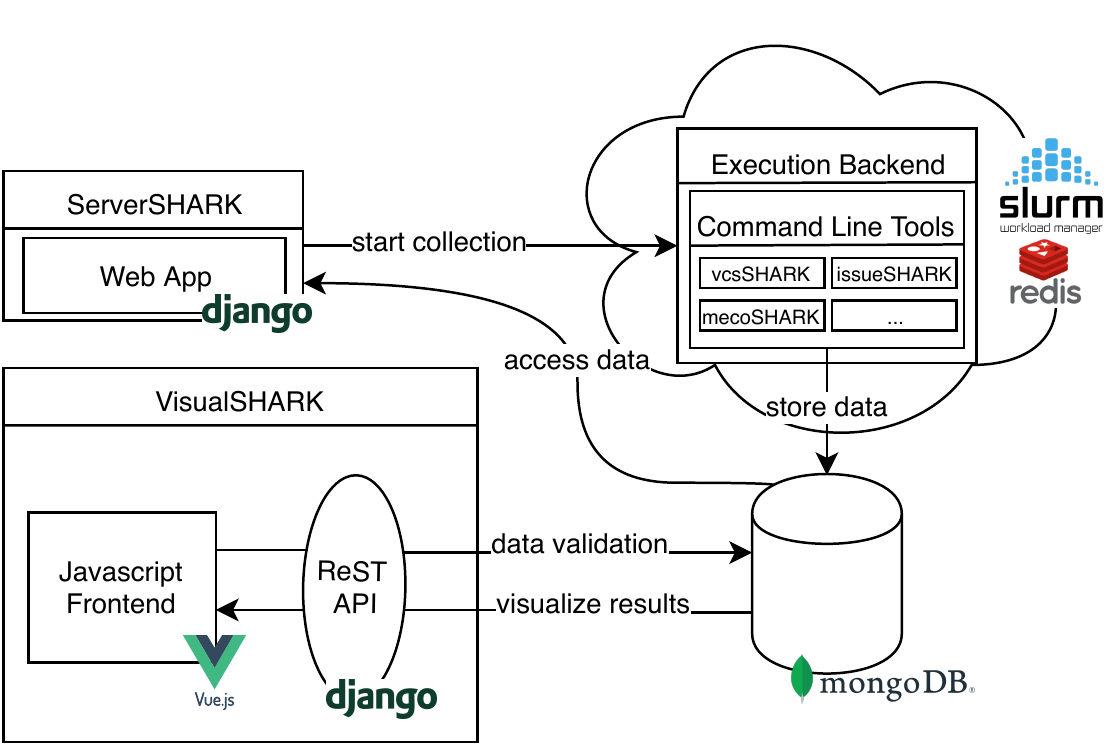}
    \caption{Overview of the SmartSHARK ecosystem}
    \label{fig:ecosystem}
\end{figure}

The SmartSHARK ecosystem is an environment for replicable and reproducible software mining research. Figure~\ref{fig:ecosystem} gives an overview of the four major components within SmartSHARK and how they interact: 1) a set of command lines tools that can be used for the data collection and enrichment; 2) a MongoDB instance for the storage of all collected data; 3) the ServerSHARK web application that can be used for convenient data collection; and 4) the VisualSHARK web application that provides an overview of the collected data and enables manual validations. In the following, we describe each part of the ecosystem in greater detail. More information, including a demonstration video can be found online\footnote{https://smartshark.github.io/\\https://youtu.be/69ongpoBtQg}.

\subsection{Command Line Tools}

The core of the SmartSHARK ecosystem is a constantly growing collection of tools that can be executed directly on the command line. These tools can be divided into tools that scrape data directly from software repositories and tools that enrich data. All tools work with a shared database (Section~\ref{sec:mongodb}).

\subsubsection{Data Collection}

Currently, there are ten tools in the Smart\-SHARK ecosystem that collect data from software repositories.
\begin{itemize}
  \item vcsSHARK: collects the development history from version control systems, currently supporting only git. 
  \item issueSHARK: collects issue tracking data, e.g., from Jira, Bugzilla, and GitHub issues.
  \item mailingSHARK: collects data from mailing list archives.
  \item travisSHARK: collects logs from the Travis CI continuous integration system.
  \item mecoSHARK: collects software metrics with the OpenStaticAnalyzer\footnote{https://github.com/sed-inf-u-szeged/OpenStaticAnalyzer}. 
  \item coastSHARK: collects software metrics about node counts in abstract syntax trees as well as the import statements from files. 
  \item readabilitySHARK: collects code readability evaluations from models by Buse\etal~\cite{Buse2010} and Scalabrino\etal~\cite{Scalabrino2018}.
  \item refSHARK: determines refactorings using RefDiff by Silva\etal~~\cite{Silva2017}.
  \item rMineSHARK: determines refactorings using RefactoringMinder by Tsantalis\etal~\cite{Tsantalis2018}. 
  \item changeSHARK: determines the type of changes to Java files with ChangeDistiller by Fluri\etal~\cite{Fluri2007} using the classification from Zhao\etal~\cite{Zhao2017}.
\end{itemize}

The collection tools automatically map fields that share the same content to the same fields in the shared database. For example, SmartSHARK's data model for issue tracking data is based on the Jira issue tracker. Fields from other issue trackers are mapped to this data model as closely as possible. In SmartSHARK, the body of the first post of a GitHub issue is the same as the description of the issue in Jira, the subsequent posts on GitHub are the same as comments in Jira. Thus, the tools of the SmartSHARK ecosystems harmonize data from different information sources and, thereby, simplify downstream analysis with heterogeneous data sources. 

The tools travisSHARK, mecoSHARK, coastSHARK, refSHARK, rMineSHARK, and changeSHARK all automatically link the collected data to the respective commits in the version control system. For example, travisSHARK extends the commit with the travis log, mecoSHARK with static software metrics for all files in the repository at the time of the commit, and rMineSHARK creates a list of all refactorings that were performed in the commit. Thus, these tools automatically link their collected information with each other using the commits, thereby enriching the data collected from the version control system. To ensure replicability of the execution of the collection tools as well as to facilitate the further extension of the data, the vcsSHARK stores an archive of the repository as a tar-ball as part of the shared database\footnote{Only possible with Git as version control system}. 

\subsubsection{Data Enrichment}

Moreover, there are tools in the SmartSHARK ecosystem that heuristically enrich the collected data. 
\begin{itemize}
    \item linkSHARK: establishes links from commits to issues, e.g., using the SZZ algorithm~\cite{Sliwerski2005} or based on matching Jira issue ids. 
    \item labelSHARK: determines labels for commits, e.g., whether a commit is bugfixing based on linked issues, whether any refactoring was performed in the commit, if code documentation was modified, or whether the authors added or removed self admitted technical debt. 
    \item inducingSHARK: determines bug inducing changes, e.g., using the SZZ algorithm.
    \item identitySHARK: identifies different identities of the same developer, e.g., due to the use of different email addresses or different spellings of their name. 
\end{itemize}

The main purpose of these tools is to establish connections between data that was collected previously and, thereby, enable a deeper analysis of the collected data. 

\subsection{MongoDB}
\label{sec:mongodb}

All command line tools use a single MongoDB database for the storage of the collected data. MongoDB\footnote{https://www.mongodb.com/} is a NoSQL database for document storage that scales well with large amounts of data and can be used both on a single machine as well as in a distributed cluster. Thus, MongoDB is suitable to store large amounts of collected data. Moreover, MongoDB uses a document-based and schema-free approach that uses JSON-like representations of the data. This enables flexible changes to the schema of the collected data, without overhead, e.g., due to database migrations to modify the schema. The use of a single database enables synergy effects between the different command line tools. For example, changes in the values of static software metrics can easily be correlated with refactorings, because both are available in the same database and linked to the commits. A documentation of the current database schema can be found online\footnote{https://smartshark2.informatik.uni-goettingen.de/documentation/}. 

There are drivers for MongoDB for many technologies that enable access to the database. Additionally, we developed the libraries pycoSHARK for Python and jSHARK for Java that provide Object-Relational Mappings (ORM) for convenient access to the data. These libraries are available on PyPi and Maven Central, respectively.

\subsection{ServerSHARK}

ServerSHARK is a Web application that simplifies the data collection with SmartSHARK. Data collection tools can be installed in ServerSHARK directly from GitHub. ServerSHARK can then trigger the collection of data with these tools. The command line tools are either executed locally with a redis queue\footnote{https://redis.io/} or remotely in a batch processing system. ServerSHARK currently supports SLURM\footnote{https://slurm.schedmd.com/}, but we have also used LSF\footnote{https://www.ibm.com/marketplace/hpc-workload-management} in the past. This batch execution is not only a convenience feature, but a key requirement for the scalable collection of static software metrics for complete software repositories. For example, the Apache Jena project has over 20.000 commits and the execution of the mecoSHARK for the collection of static metrics requires about 30 minutes of time per commit. Thus, roughly 416 days of computational time are required for the collection of this data. In a batch system with hundreds of compute nodes, this task can be solved in a matter of hours.

The drawback of the distributed execution of data collection in batch processing systems is that these systems tend to be unreliable, i.e., it is quite likely that single jobs fail in case thousands of jobs are executed. The reason for this is that issues like hardware failures or network problems are much more likely to happen if hundreds of nodes are involved in comparison to a single machine. To account for this, ServerSHARK can check the execution logs of the batch system to detect failures. Since some failures are silent (sometimes jobs just vanish), ServerSHARK can also actively check the consistency between the clone of a repository and the collected data, e.g., if metric data is available for each file that exists in a revision of the source code.

\subsection{VisualSHARK}

VisualSHARK is a Web application that provides a dashboard that gives insights into aspects of the collected data. VisualSHARK provides basic statistics about the collected data, e.g., the number of issues or files per project. For each project, it is possible to browse the commit history and the issues and to inspect the links that were established. Moreover, VisualSHARK can display a commit graph that can, e.g., show all commits, only bug fixing commits, or all commits where the messages matches a certain query. This enables the manual inspection of the collection data. 

\begin{figure}
    \centering
    \includegraphics[width=\linewidth]{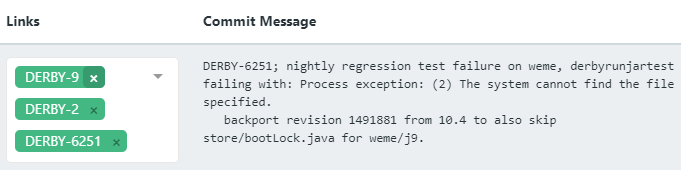}
    \caption{Validation of links between commits and issues with the VisualSHARK. The VisualSHARK suggests the links that heuristics like SZZ detect and researchers can remove invalid links, i.e., DERBY-2 and DERBY-9 in the example.}
    \label{fig:linkvalidation}
\end{figure}

\begin{figure}
    \centering
    \includegraphics[width=\linewidth]{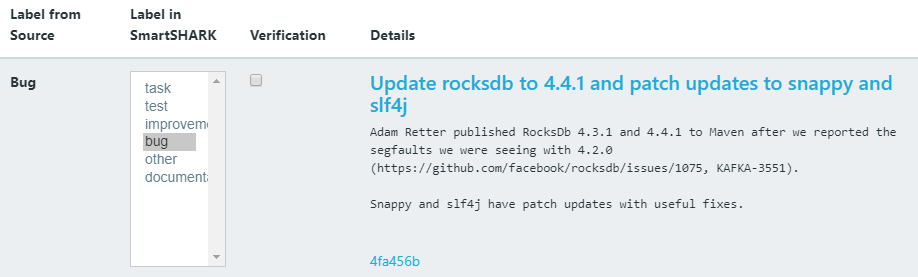}
    \caption{Validation of the developer assigned issue types. Researchers can select the actual type of the issue hand have to check that they have validated the type. They can easily access further information by clicking the title of the issue to access the original issue in the issue tracker or linked commits by clicking on the revision hash.}
    \label{fig:issuevalidation}
\end{figure}

\begin{figure}
    \centering
    \includegraphics[width=\linewidth]{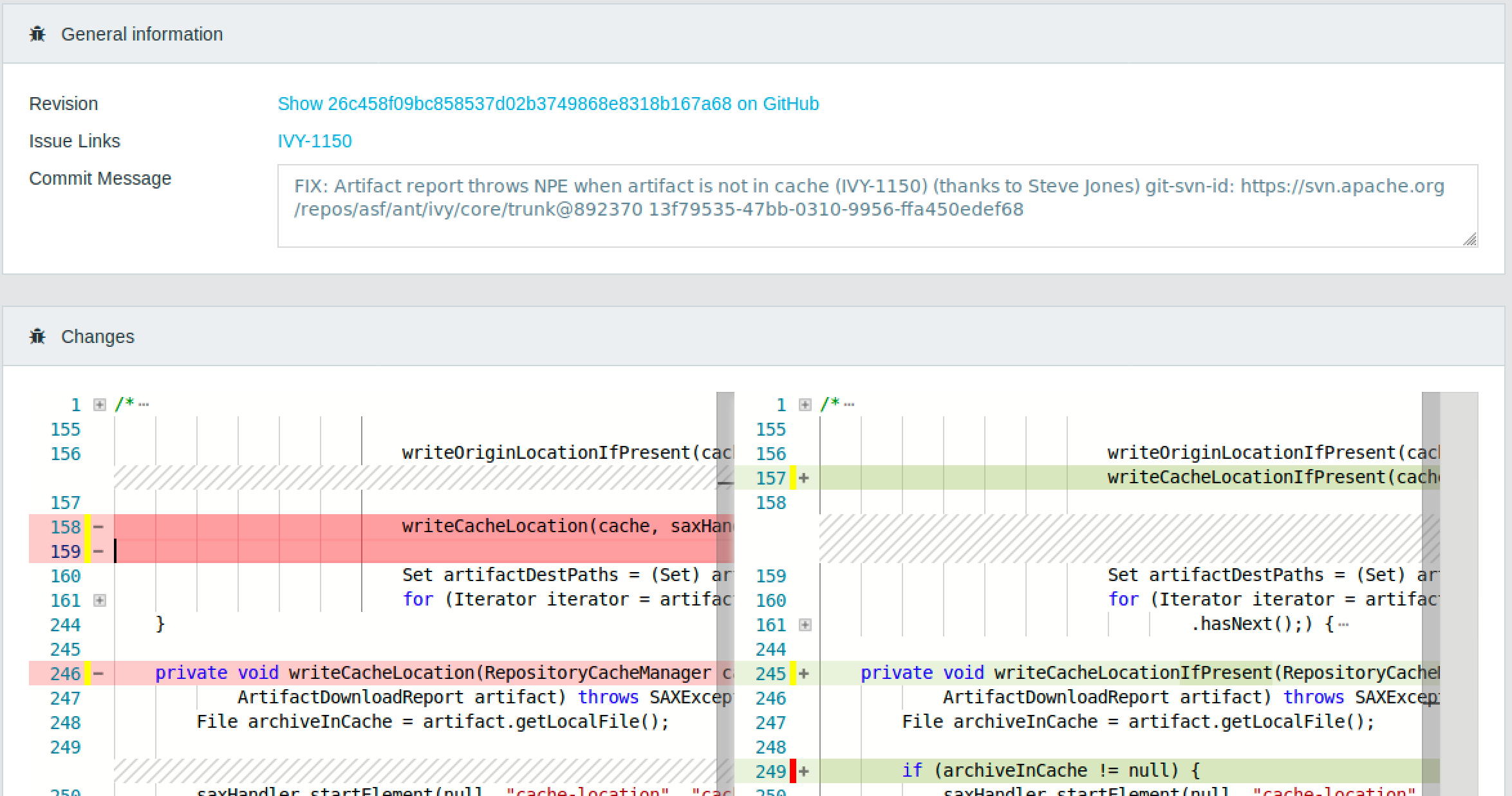}
    \caption{Validation of the lines that actually contributed to the logical change required to fix a defect. The markers at the beginning of the lines are added by the validator. Red indicates part of the bugfix, yellow refactoring. }
    \label{fig:hunkvalidation}
\end{figure}

Moreover, VisualSHARK also supports the manual validation of data and the storage of the validation results in the MongoDB. Currently, VisualSHARK allows two manual validations: each link that is established from a commit to an issue can be manually determined as correct or incorrect (Figure~\ref{fig:linkvalidation}). Similarly, we allow the validation of the types of issues similar to the work by Herzig\etal~\cite{Herzig2013}. For this, the researchers get the developer classification of the issue, the title and the description of the issue, a link to the actual issue in the projects issue tracker, as well as links to all commits that were linked to an issue (Figure~\ref{fig:issuevalidation}). A manual validation of which lines in a change actually contributed to bug fixes based on Visual Studio Code's editor\footnote{https://code.visualstudio.com/} is currently being added to VisualSHARK (Figure~\ref{fig:hunkvalidation}). 

\section{Analyzing Data with SmartSHARK}
\label{sec:analysis}

We already used data collected with SmartSHARK for multiple publications, e.g., on differences between unit and integration tests~\cite{Trautsch2019}, the impact on static analysis~\cite{Trautsch2019a, Trautsch2019b}, the mining of project activity patterns~\cite{Herbold2019a}, or the detailed analysis of issues with defect prediction data~\cite{Herbold2019}. As part of the latter, we developed the Python script Mynbou\footnote{https://github.com/smartshark/mynbou}, that can collect release level data for defect prediction. We use Mynbou as a showcase for the capabilities of SmartSHARK, because it uses large parts of the SmartSHARK database, including the links that the different tools establish between the data. Mynbou uses inducing changes determined by inducingSHARK for bug fix labels based on manually validated data with the VisualSHARK. Mynbou computes features from the source code history made available by vcsSHARK and metrics computed by coastSHARK and mecoSHARK. These are fairly standard for defect prediction data, albeit not in the scale and diversity provided by Mynbou. Mynbou can also easily enhance the defect prediction data with features that were not considered in the state of the art before, e.g., data about the number of refactorings, because such data is also readily available in the SmartSHARK database. 

\section{Related Work}
\label{sec:related-work}

There are too many tools that consider mining of software repositories to conduct a full review here. Instead, we focus on two very popular and powerful tools that are currently often used for similar purposes as SmartSHARK: GrimoireLab\footnote{https://chaoss.github.io/grimoirelab/} and PyDriller~\cite{Spadini2018}. In comparison to other popular tools like GHTorrent~\cite{Gousios2013}, these tools also process data and do not only scrape them from repositories. 

The SmartSHARK ecosystem shares many properties with \linebreak GrimoireLab. GrimoireLab also provides command line tools for data collection, a server that can trigger the execution of the command line tools, a single backend for storing data, and a front-end that can, e.g., provide a dashboard for analysis. However, there are several notable and important differences between GrimoireLab and SmartSHARK. In general, GrimoireLab is more powerful with respect to the number of different data sources that are supported. However, the command line tools from GrimoireLab do not enforce a common schema for different data sources of the same type. Instead, all data stays in the schema it was retrieved, i.e., Jira issues have a different schema than GitHub issues. While the command line tools can also write into a shared database, this database also does not enforce a common schema. Instead, GrimoireLab uses an ElasticSearch database for the storage of all data and requires users of this database to cope with the diversity of the data. In comparison, all tools in the SmartSHARK ecosystem share the same underlying data representation and, thereby, provide compatible representations for downstream analysis. 

The second difference is the depth of the analysis. While the number of data sources and the ability to visualize the collected data in dashboards is the strength of GrimoireLab, the data that GrimoireLab gathers from each data source is limited. Moreover, there is only little linking between data from different sources, except via the authors. For example, while GrimoireLab collects the commit messages, these messages are not used to establish links to the collected issue tracking data. Moreover, GrimoireLab does not analyze source code directly, e.g., to collect software metrics. While the tools in the SmartSHARK ecosystem support fewer data sources, they implement detailed data collection and enrichment methods for the available data.

A second notable tool that has similarities to parts of the Smart\-SHARK ecosystem is PyDriller~\cite{Spadini2018}. PyDriller is a very fast and powerful tool to scrape the development history of Git projects. \linebreak PyDriller even allows the collection of a small set of software metrics for each revision, as well as the identification of bug fixing and bug inducing commits. Moreover, PyDriller is very lightweight and can easily be used. The main drawback of PyDriller in comparison to SmartSHARK is the limitation on Git repositories as data sources. This limits, to some degree, the capabilities of PyDriller, because, e.g., the identification of bug fixing commits cannot utilize data from an issue tracker. Moreover, the amount of data for each commit that SmartSHARK can collect for each release is larger than for PyDriller. However, data collection with SmartSHARK is more complex and requires more computational resources.

\section{Conclusion}
\label{sec:conclusion}
SmartSHARK is a versatile and still growing ecosystem for software repository mining that facilitates both the replication and comparison of existing work, as well as the development of new analysis approaches on software repository data. 

\begin{acks}
This work is partially funded by DFG Grant 402774445. We also thank the GWDG for their great support while we were using their batch system. 
\end{acks}

\bibliographystyle{ACM-Reference-Format}
\bibliography{literature}

\end{document}